\documentclass[useAMS,usenatbib]{mn2e}
\usepackage{graphicx}
\usepackage{latexsym}
\usepackage{psfig}
\usepackage{amssymb,amsmath}
\usepackage{subfig}
\usepackage{caption}
\usepackage{epstopdf}
\usepackage{float}

\title{Turbulence in Giant Molecular Clouds: The effect of photoionisation feedback}

\author[D. M. Boneberg, J. E. Dale,  P. Girichidis, B. Ercolano]{D. M. Boneberg$^{1}$\thanks{E-mail: boneberg@usm.lmu.de }, J. E. Dale$^{1,2}$, P. Girichidis$^{3}$, B. Ercolano$^{1,2}$\\
$^{1}$Universit\"{a}ts Sternwarte M\"{u}nchen, Scheinerstr. 1, 81679 M\"{u}nchen, Germany.\\
$^{2}$Excellence Cluster `Universe', Boltzmannstr. 2, 85748 Garching, Germany.\\
$^{3}$Max Planck Institut f\"{u}r Astrophysik, Karl-Schwarzschild-Str. 1, 85741 Garching, Germany.}

\begin{document}

\pagerange{\pageref{firstpage}--\pageref{lastpage}} \pubyear{2014}

\maketitle

\label{firstpage}

\def\mnras{MNRAS}
\def\apj{ApJ}
\def\aj{AJ}
\def\aap{A\&A}
\def\apjl{ApJL}
\def\apjs{ApJS}
\def\araa{ARA\&A}
\def\pasp{PASP}
\def\pre{Phys.~Rev.~E}
\def\aapr{A\&ARv}
 
\begin{abstract}
Giant Molecular Clouds (GMCs) are observed to be turbulent, but theory shows that without a driving mechanism turbulence should quickly decay. The question arises by which mechanisms turbulence is driven or sustained. It has been shown that photoionising feedback from massive stars has an impact on the surrounding GMC and can for example create vast HII bubbles. We therefore address the question of whether turbulence is a consequence of this effect of feedback on the cloud. To investigate this, we analyse the velocity field of simulations of high mass star forming regions by studying velocity structure functions and power spectra.
We find that clouds whose morphology is strongly affected by photoionising feedback also show evidence of driving of turbulence by preserving or recovering a Kolmogorov-type velocity field. On the contrary, control run simulations without photoionising feedback have a velocity distribution that bears the signature of gravitational collapse and of the dissipation of energy, where the initial Kolmogorov-type structure function is erased.
\end{abstract}

\section{Introduction}
\label{sec:Intro}
Observations of large non-thermal linewidths within molecular clouds are interpreted as an indication of supersonic turbulence (e.g. \cite{ZuckermanPalmer1974, FalgaroneEtAl1994, VazquezSemadeni2000, OssenkopfMacLow2002, HeyerBrunt2004, MacLowKlessen2004, Klessen2011,Roman-DuvalEtAl2011, Dobbsetal2013}). Turbulence has an important effect on the process of star formation as it influences the stability of GMCs, which are the birthplaces of stars. Depending on the scales on which it is acting, turbulence can either prevent gravitational collapse (large scales) or trigger it (small scales) \citep{MacLowKlessen2004}. 

The characteristic distribution of densities in molecular clouds can be explained by turbulent motions. The width of the probability density function (PDF) of the gas density depends on the Mach number of the turbulence \citep{Vazquez94,PassotVazquezSemadeni1998, FederrathEtAl2008, PriceEtAl2011, MolinaEtAl2012, KonstandinEtAl2012}. Stronger turbulence thus leads to higher densities and can trigger gravitational collapse. The density PDF and its dependence on turbulence properties has served as a basis for star formation theories in theoretical studies \citep{krumholzMcKee2005, HennebelleChabrier2008, PadoanNordlund2011, FederrathKlessen2012, PadoanEtAl2013, GirichidisEtAl2014} as well as an observational tool to determine the dynamical state of molecular clouds \citep{KainulainenEtAl2009, BruntEtAl2010, BallesterosParedesetal2011a, SchneiderEtAl2012, SchneiderEtAl2013, KainulainenTan2013, KainulainenEtAl2014}.
 
The variety of complex astrophysical motions interpreted as turbulence can be quantified by a spectral energy cascade and the resulting spectral energy distribution. The seminal work by \cite{Kolmogorov1941} derives the energy cascade from large to small scales as a result of eddies breaking up into smaller and smaller structures. The subsequent energy transport through all spatial scales is violated if considering highly supersonic flows in which the medium reacts compressively (Burgers turbulence). In this limit, which is the dominant limit in typical ISM and GMC environments, shocks dominate and allow the energy to be transferred across larger spatial ranges. The spectral energy distribution is steeper compared to the subsonic Kolmogorov-type turbulence \citep{KritsukEtAl2007, SchmidtEtAl2009, FederrathEtAl2009, FederrathEtAl2010, FederrathKlessen2013, Federrath2013}. However, a full theory of compressible turbulence is still missing. We will briefly discuss the implications of Kolmogorov-type and Burgers turbulence in Sections~\ref{subsec:Velocityfield} and \ref{subsec:Spectra}.
 
\cite{MacLowetal1998} show that without a driving mechanism (i.e. energy input), turbulence will decay very quickly (on the order of the crossing time of the size of the flow). Numerous driving mechanisms for turbulence, both external and internal to the GMCs have been proposed and are described in a number of dedicated reviews, including  \cite{MacLowKlessen2004, ElmegreenScalo2004, McKeeOstriker2007, Klessen2011} and \cite{Dobbsetal2013}.

There are various mechanisms by which driving on large scales could be possible. One way is to inject energy by external accretion flows, i.e. from a flow outside the GMC \citep{KlessenHennebelle2010}. Gravity-driven turbulence has also been studied by \cite{FederrathEtAl2011}. Furthermore, it is possible that the formation process of the molecular cloud itself could explain the origin and driving of turbulence. Density waves within galactic spiral arms can drive convergent flows of atomic gas \citep{VazquezSemadenietal2007, Heitschetal2008, Hennebelleetal2008}. Simulations performed by these authors confirm that GMCs can be formed from diffuse gas, which build up at the stagnation points of these large-scale flows \citep{BallesterosParedesetal2005, BallesterosParedesetal2011b}. 
Numerical simulations indeed show that ram-pressure confined flows can sustain turbulence, see e.g. \cite{VazquezSemadenietal2006}. \cite{Dobbsetal2011a, Dobbsetal2011b, Dobbsetal2013} and \cite{DobbsPringle2013} discuss galactic flows as a mechanism to inject energy at large scales, for example due to the interaction with density waves in spiral arms or cloud-cloud collisions \citep{TaskerTan2009}. 

Turbulence might also be driven by large-scale processes internal to the cloud. \cite{Krumholzetal2006} and \cite{Goldbaumetal2011} perform semi-analytic models of GMCs and conclude that expanding HII regions do have a significant effect on the velocity field as the energy content of HII regions is similar to that of the velocity field. Numerical simulations by \cite{Gritschnederetal2009} show that externally photoionising a turbulent box sustains turbulence. \cite{WalchWhitworthEtAl2012} find that photoionising radiation injects substantial amounts of thermal energy in the gas and that it is thus a potential driver of turbulence.
 
\cite{MacLowKlessen2004} discuss supernovae as a driving source. On galactic scales of the ISM, SNe seem to be indeed the main driver of turbulence, which can explain the multiphase structure of the ISM. Taking into account just disk instabilities does not offer such an explanation (see e.g. \cite{McKeeOstriker2007, HennebelleFalgarone2012}).
However, \cite{Dobbsetal2013} claim that supernovae are most likely not a very important driving mechanism on GMC scales because the timescale of stellar evolution (of very massive stars) and crossing time are of the same order of magnitude. Thus, supernovae are probably more important as a mechanism for dispersing GMCs.

Radiation pressure is a potential driving mechanism acting on smaller scales. However, \cite{KrumholzThompson2012, KrumholzThompson2013} conclude from their simulations that this mechanism is not likely to have a major effect on scales of molecular clouds.

Another feedback mechanism that could potentially drive turbulence are outflows and stellar winds. Simulations of GMCs including winds have for example been performed by \cite{RogersPittard2013}, but they do not analyse their simulations from the point of view of turbulence. 
\cite{LiNakamura2006, Wangetal2010, CunninghamEtAl2011, MyersEtAl2014, OffnerArce2014} and \cite{FederrathEtAl2014} have studied the effect of outflows, but they find that these do not play an important role on scales of the size of the cloud \citep{MacLowKlessen2004}.

In this paper, we will study the effect of photoionising feedback and the resulting HII bubbles on the turbulent velocity field within simulations of GMCs.
In Section \ref{sec:MethodSimulations} we introduce the simulations from \cite{Dale:2012tp} and describe the analysis tools used to study the velocity field of the respective molecular cloud, namely structure functions and power spectra. We continue with our findings in Section~\ref{sec:Effect} which are then also discussed in more detail in this section. Our conclusions can be found in Section \ref{sec:Conclusion}.
\section{Method and simulations used}
\label{sec:MethodSimulations}
\subsection{Simulations}
\label{subsec:Simulations}

\begin{figure*}
\centering
\begin{minipage}[c]{\textwidth}
\includegraphics[width=\textwidth]{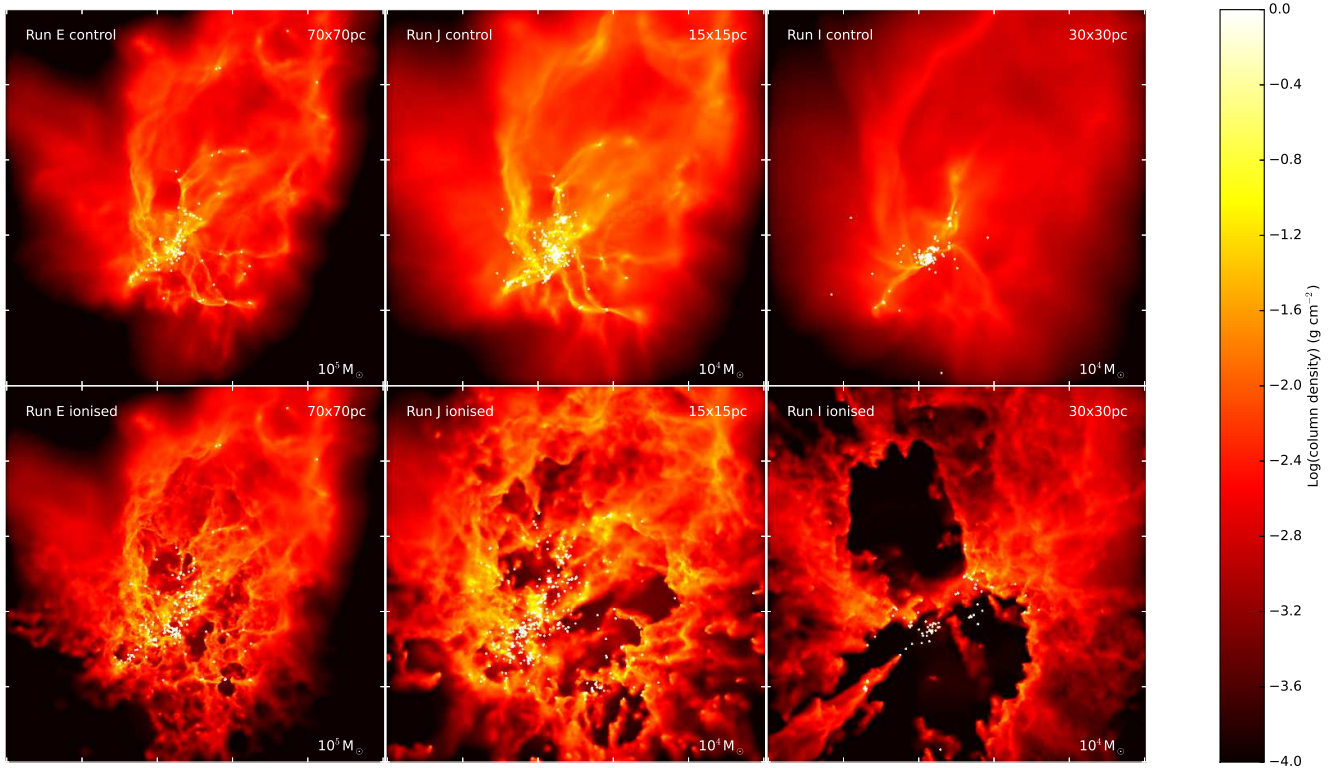}
\end{minipage}
\caption{Final snapshots of clouds (column density maps), from left to right: Run E ($t=7.74\text{Myr}$), Run J ($t=3.49\text{Myr}$) and Run I ($t=7.58\text{Myr}$). The upper row shows the control simulations, the lower one the runs including photoionisation. White dots represent sink particles, they are not to scale. Note that the size scales are varying between the plots.}
\label{fig:snapshots}
\end{figure*}

The simulations studied originate from \cite{Dale:2012tp}, we will use their nomenclature for the respective clouds. They use a hybrid N-body SPH code to simulate GMCs with initial masses of $M=10^4, 10^5$ and $10^6{\text M}_\odot$ and initial radii ranging from $5$ to $180\text{pc}$. The clouds are all seeded with turbulence.
The imposed velocity field has a power spectrum $P(k)\propto k^{-4}$ appropriate for Burgers turbulence and is generated in Fourier space with modes between k=4 and k=128 populated, before being transformed into real space. The normalisation of the velocity field is adjusted to give the clouds the desired virial ratio between gravitational and turbulent kinetic energy. Turbulence is not driven artificially in the simulations. The initial velocity field contains a ratio of compressive to solenoidal modes of 1:2. The power in the respective modes will be explained and studied in more detail in Sections~\ref{subsec:Spectra} and \ref{subsec:Modes}. The importance of this ratio for the star formation rate and morphology of GMCs is discussed in \cite{GirichidisEtAl2011}.

Self-gravitational forces between gas particles are included and calculated using a binary tree. Gravitational forces between sink and gas particles are computed by direct summation and sink particles may accrete gas particles and thus grow in mass. 
The clouds are initially gravitationally bound, with $E_\text{kin}/|E_\text{pot}|=0.7$, but they are unconfined and thus the outer regions can expand in all runs. 
\cite{Dale:2012tp} allowed the clouds to evolve and form stars, which are modelled in the simulations by sink particles: In their $10^{4}$M$_{\odot}$ clouds (including Runs I and J, which we examine here), the sink particle radius is taken to be $5\times10^{-3}$pc, the sink particle formation density is then $7\times10^{7}$cm$^{-3}$ and the sinks in these simulations are treated as individual stars. Once three sinks have grown to masses exceeding $20$M$_{\odot}$, photoionisation is enabled in these calculations.\\
\indent In the more massive clouds, sink particles represent small stellar subclusters. Here, we use results from Run E, a $10^{5}$M$_{\odot}$ cloud, in which the sink particle radius is 0.1pc and the sink formation density is $4\times10^{5}$cm$^{-3}$. To determine whether a given subcluster sink may be a source of ionising photons, the subcluster is assumed to have a Salpeter mass function in the range $0.1$ to $100\text{M}_{\odot}$. The mass in stars of more than 30M$_{\odot}$ is computed and divided by 30M$_{\odot}$ and rounded to the nearest integer. The subcluster is then assigned a flux corresponding to this integral number of 30M$_{\odot}$ stars. Photoionisation is enabled in the $10^{5}$M$_{\odot}$ clouds once three such subclusters have formed.

\cite{Dale:2012tp} use the photoionisation code from \cite{Daleetal2007} and \cite{DaleBonnell2011}. The code uses a simple ray-tracing algorithm and a Str{\"o}mgren volume technique to compute the flux of ionising photons arriving at a given SPH particle and update its ionisation state accordingly. The ionisation algorithm was modified in a simple way in \cite{DaleBonnell2011} to take into account the effect of multiple ionising sources with overlapping HII regions.

In order to isolate the impact of photoionisation radiation, control simulations were performed. The setup for these clouds is identical to the one just described, the only difference is that photoionisation feedback is absent there.

\cite{Dale:2012tp} find that their most massive clouds are hardly affected by photoionising radiation due to their high escape velocities. Contrarily, in the clouds with less mass and lower density, ionising feedback creates huge HII bubbles. This can be seen in the final snapshots of three representative simulations in Figure \ref{fig:snapshots}. The simulations are stopped before the first supernovae are expected to explode. Due to the different cloud properties, they need different time spans until the first stars are born and thus the final times of the simulations vary accordingly (Run~E: $t_\text{f}=\nolinebreak 7.74\text{Myr}$, Run~J: $t_\text{f}=3.49\text{Myr}$, Run~I: $t_\text{f}=7.58\text{Myr}$). The properties of these clouds are listed in Table~\ref{tab:properties}.
The upper row of Figure~\ref{fig:snapshots} displays the final snapshots of the control runs of Run E, J and I, the lower row those of the corresponding runs including photoionisation.

Run E, the cloud that is displayed on the left, has an initial mass of $M_\text{ini}=10^5\text{M}_\odot$ and an initial radius of $r_\text{ini}=\nolinebreak21\text{pc}$. In this case, ionisation does not have a big impact on the star formation rate, but it has some effect on its density distribution as photoionising feedback is beginning to open some small HII bubbles. This is different in the cloud of Run J that has $M_\text{ini}=10^4\text{M}_\odot$ and $r_\text{ini}=5\text{pc}$ (Fig. \ref{fig:snapshots}, centre). The morphology is changed substantially by feedback which creates a complex system of HII bubbles and pillar-like features. In Run I (with $M_\text{ini}=10^4\text{M}_\odot$ and $r_\text{ini}=10\text{pc}$), the effect of feedback on the cloud morphology is even greater: two HII bubbles are opened  up, which occupy a very large fraction of the cloud volume, in addition to some pillar-like structures. 

We have studied the velocity structure functions of 13 molecular clouds, including also unbound ones taken from \cite{Daleetal2012unbound} (where $E_\text{kin}/|E_\text{pot}|=2.3$, see Section~\ref{subsec:Unbound}). Here we choose three representative, bound clouds, namely Run E, J and I, to illustrate the behaviour we found.
\begin{center}
\begin{table*}
\begin{tabular}{c c c c c c c c c c c c c }
Run & $M_{\text{ini}}$ $(\text{M}_\odot)$ & $r_{\text{ini}}$ $(\text{pc})$  & $v_\text{RMS}$ $\left(\frac{\text{km}}{\text{s}}\right)$  & $\tau_\text{cr}$ $(\text{Myr})$ & $t_\text{f}$ $(\text{Myr})$ &

 $t_\text{f}$ $(\tau_\text{cr})$ &

 $r_\text{s}$ $(\text{pc})$ & $n_\text{s}$ $\left(\frac{1}{\text{cm}^{3}}\right)$ & $\frac{E_\text{kin}}{|E_\text{pot}|}$ & $\mathcal{M}_\text{ini}$ &  $\mathcal{M}_\text{ion}$ \\ \hline \hline
E & $10^5$ & $21$ & $4.6$ & $4.47$ & $ 7.74$& $1.73$ & $0.1$ & $4\times 10^5$ & 0.7 &23.0 & 14.5 \\ \hline
J & $10^4$ & $10$ & $3.0$ & $1.63$ & $3.49$ & $2.14$ & $5\times 10^{-3}$ & $7\times 10^7$ & 0.7 & 15.0 & 9.0 \\ \hline
I & $10^4$ & $5$ & $2.1$ & $4.66$ & $7.58$ & $1.63$ & $5\times 10^{-3}$& $7\times 10^7$ & 0.7 & 10.5 & 7.0\\
\hline
\end{tabular}
\caption{Properties of the different clouds: initial mass, initial radius, initial turbulent velocity dispersion, initial crossing time, final time of simulation, final time of simulation divided by the initial crossing time, sink particle radius, sink formation density, ratio of kinetic to potential energy, initial turbulent Mach number, Mach number at start of ionisation}
\label{tab:properties}
\end{table*}
\end{center}
\subsection{Characterising the velocity field using structure functions}
\label{subsec:Velocityfield}
We analyse the effect of photoionising radiation on the gas velocity field of the surrounding cloud using two different approaches, namely \textit{velocity structure functions} and \textit{power spectra}. For the first approach we use velocity structure functions of second order 
\begin{equation}
S_2({dr})= \langle |\vec{v}(\vec{x}) - \vec{v}(\vec{x}+\vec{dr})|^2 \rangle\equiv\langle \delta v^2 \rangle \hspace{2pt}.
\end{equation}
Due to symmetry, the structure function depends only on the absolute value of the separation $\vec{dr}$ and not on its direction.
In the case of an incompressible turbulent fluid, the structure function will be 
\begin{equation}
S_2({dr}) \propto dr^{\frac{2}{3}} \hspace{2pt};
\end{equation}
see the review by \cite{ElmegreenScalo2004} for a detailed description and derivation of structure functions. We restrict ourselves to structure functions of second order here (corresponding to  a proportionality to the velocity squared). Thus we omit the index $2$ from here on and refer to the structure functions as $S(dr)$.

In Section~\ref{subsec:Simulations} we described the initial seeding of the molecular clouds with turbulence. The underlying power spectrum is such that the structure functions of all the clouds are initially of power law shape. This is visible in the plots in Section~\ref{subsec:timeevol} and is analysed in more detail there.

\begin{figure}
\centering
\includegraphics[width=0.8\columnwidth]{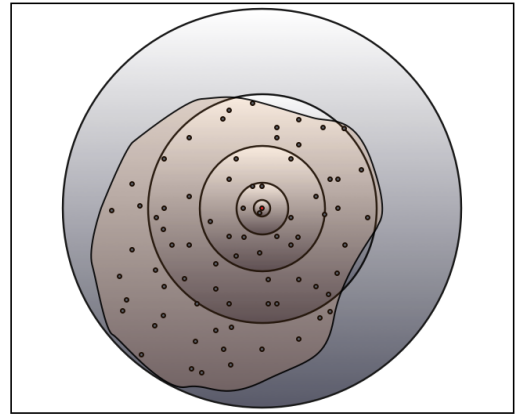}
\caption{Schematic illustration of binning process. The brownish feature represents the molecular cloud, the dots are the SPH particles. The sample particle is illustrated by the central dot and the circles indicate the logarithmically spaced radial bins.}
\label{fig:binninglog}
\end{figure}

The calculation of the structure functions is done in the following way:
We use $10^4$ randomly-chosen sample particles $j$, out of the initially $10^6$ particles, around which we radially bin the other particles $i$. The bins are logarithmic as the dynamic range of scales is large. We have checked that increasing the number of sample particles by a factor of three does not change the resulting structure functions, we therefore restrict ourselves to $10^4$ particles to keep the computing time limited. We then calculate the mean of the square of the velocity difference between the sample particle and all other particles. Then we average the velocity differences in the respective bins and repeat the procedure for the other sample particles:
\begin{equation}
S({dr})=\langle \left( \vec{v_i} - \vec{v_j} \right)^2 \rangle _{bin} = \langle \delta v^2 \rangle _{bin} \hspace{2pt}.
\end{equation}
This is schematically illustrated in Figure~\ref{fig:binninglog}, where the brown, irregular shape is the molecular cloud. The red dot in the middle represents the randomly chosen sample particle, around which all other SPH particles (dark dots) are put into logarithmically spaced radial bins. 

In addition, we have performed a test for the control simulation of Run I to determine the influence of higher resolution on the structure functions. We find that increasing the initial number of SPH particles to $10^7$ does not significantly alter the resulting $S(dr)$, except at scales $\lesssim 0.1\text{pc}$, which is approaching the sink radius in the low-resolution runs.

\subsection{Characterising the velocity field using spectra}
\label{subsec:Spectra}
\cite{Kolmogorov1941} describes an energy cascade, where eddies of decreasing size transport energy from the large scale (where it is injected) to the small scales (where energy is dissipated due to friction). This description is only valid for incompressible fluids and the resulting velocity power spectrum, 
\begin{equation}
  P(k)dk=4\pi k^2 \hat{\mathbf{v}}(k)\cdot \hat{\mathbf{v}}(k)dk \hspace{2pt},
\end{equation}
shows a scaling of $v^2$, which is equal to the energy $E$ for incompressible fluids  \citep{ElmegreenScalo2004}
\begin{equation}
E(k)\propto k^{-5/3} \hspace{2pt}.
\end{equation}
We note that this scaling is for the one-dimensional power spectrum assuming isotropy in Fourier space ($E(|\mathbf{k}|) =\nolinebreak E(k)$).
Molecular clouds can in general not be considered an incompressible fluid. In compressible fluids dominated by shocks (Burgers turbulence), the energy spectrum will be
\begin{equation}
E(k)\propto k^{-2} \hspace{2pt}.
\end{equation}
Nevertheless, \cite{FederrathEtAl2009,FederrathEtAl2010} find that the exponent of the energy spectrum of both observations and simulations will be between $-5/3$ and $-2$, i.e. the values lie between the cases for incompressible and shock-dominated turbulence (for Mach numbers between $5-6$). This is consistent with the findings by \cite{KritsukEtAl2007} and \cite{SchmidtEtAl2009}. The exponent of the energy spectrum asymptotically approaches the Burgers limit at very high Mach numbers as shown in \cite{Federrath2013}.
In addition, the clouds we are studying are not isothermal, thus the Kolmogorov-type turbulent energy cascade is only an approximation.

Turbulent flows in compressible media are statistically composed of compressive (curl-free) and rotational (divergence-free) modes. In fully developed isothermal turbulence in three dimensions, the statistical ratio of compressive to solenoidal modes is 1:2 (see e.g. \cite{FederrathEtAl2008}). In order to analyse the impact of ionisation feedback with a focus on the driving mode, we investigate the compressive and solenoidal contribution to the velocity field. We transform the velocity field, $\mathbf{v}(\mathbf{x})$, into Fourier space, $\mathbf{\hat{v}}(\mathbf{k})$, and project the motions into compressive and solenoidal parts with the operators
\begin{equation}
  \mathcal{P}_\perp = \delta_{ij} - k_ik_j/k^2\quad\mathrm{and}\quad\mathcal{P}_\parallel = k_ik_j/k^2 \hspace{2pt},
\end{equation}
where $i,j\in\,x,y,z$. We then compare the ratio of both components as a function of scale using isotropic power spectra. We calculate mass-weighted spectra where we transform $\rho v^2$ in the cubic box with equally sized cells.

\section{Effect of photoionisation feedback}
\label{sec:Effect}
\subsection{Density and velocity PDFs}
\label{sec:PDFs}

\begin{figure*}
\centering
\begin{minipage}[c]{0.3\textwidth}
\centering
\includegraphics[width=\textwidth]{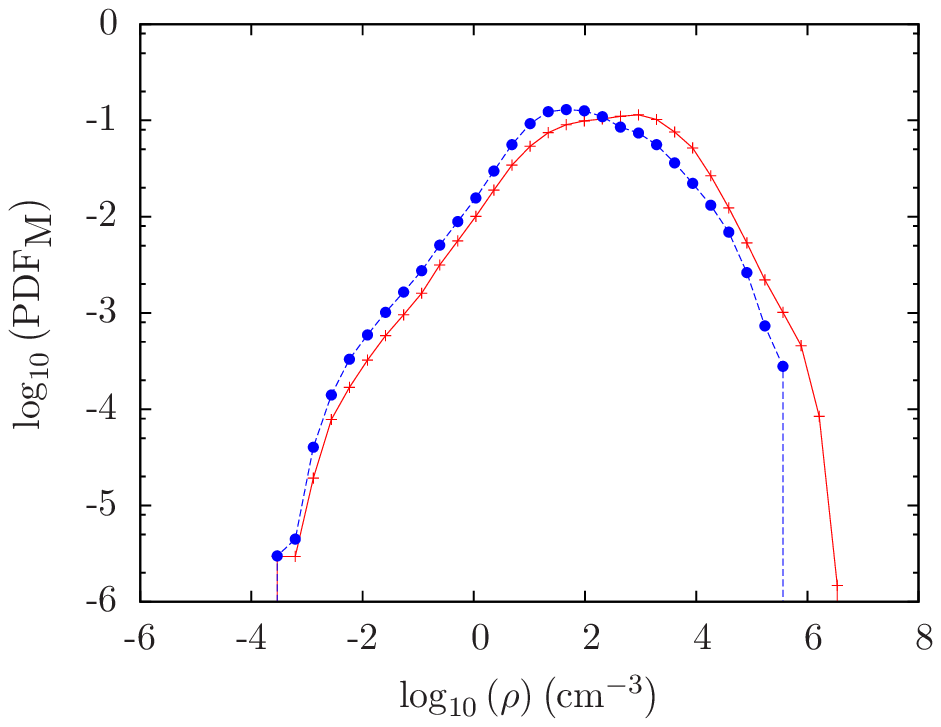}
\\
{Run E}
\label{fig:DensPDFRUNE191}
\end{minipage}
\hfill
\begin{minipage}[c]{0.3\textwidth}
\centering
\includegraphics[width=\textwidth]{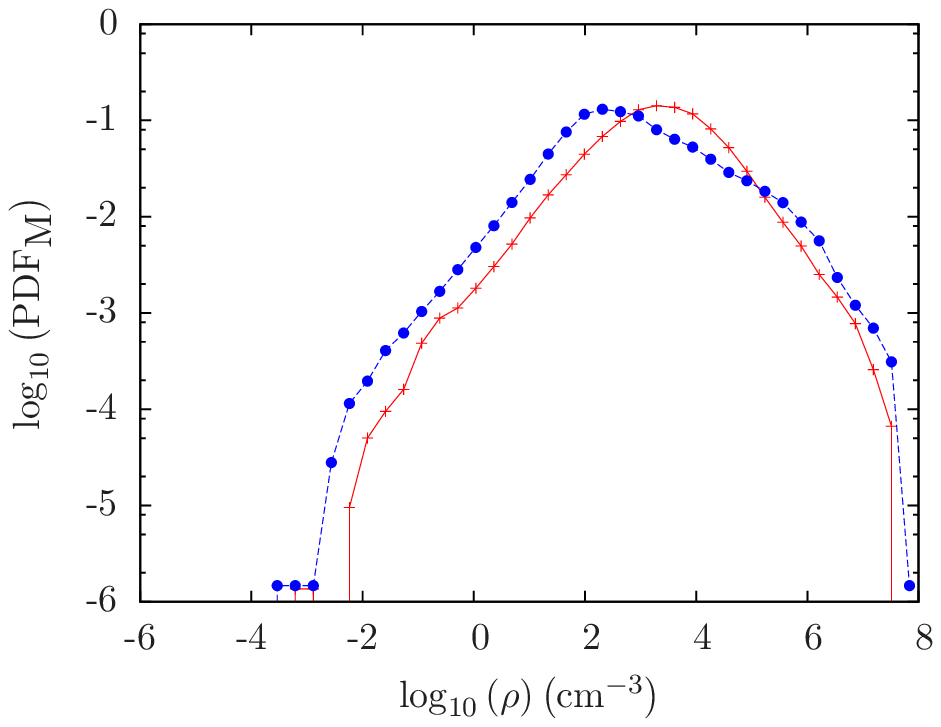}
\\
{Run J}
\label{fig:DensPDFRUNJ113}
\end{minipage}
\hfill
\begin{minipage}[c]{0.3\textwidth}
\centering
\includegraphics[width=\textwidth]{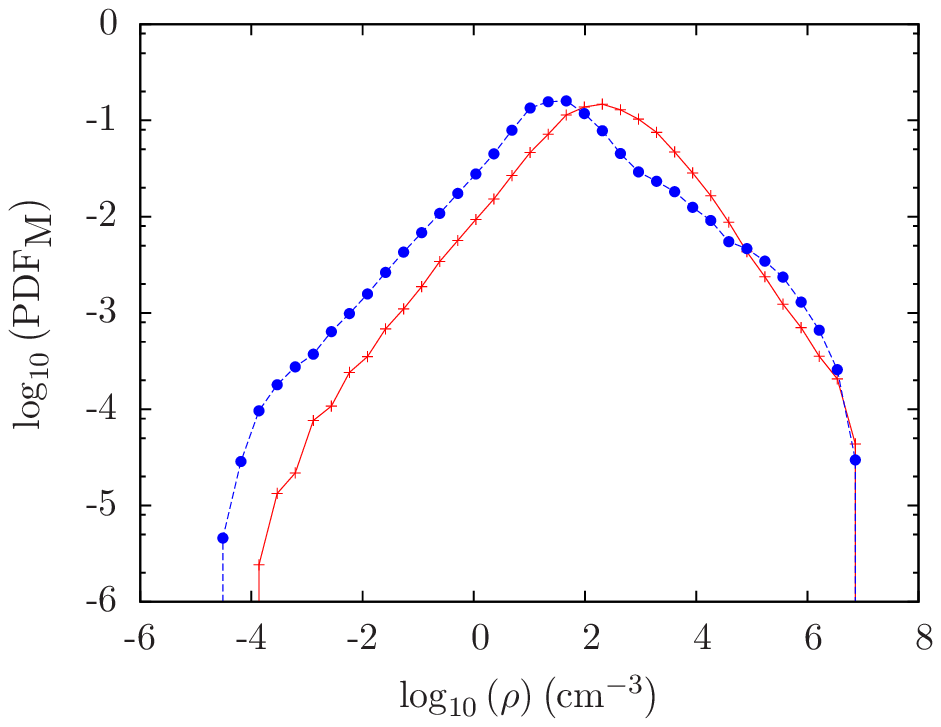}
\\
{Run I}
\label{fig:DensPDFRUNI180}
\end{minipage}
\vspace{15pt}\\
\centering
\begin{minipage}[c]{0.3\textwidth}
\centering
\includegraphics[width=\textwidth]{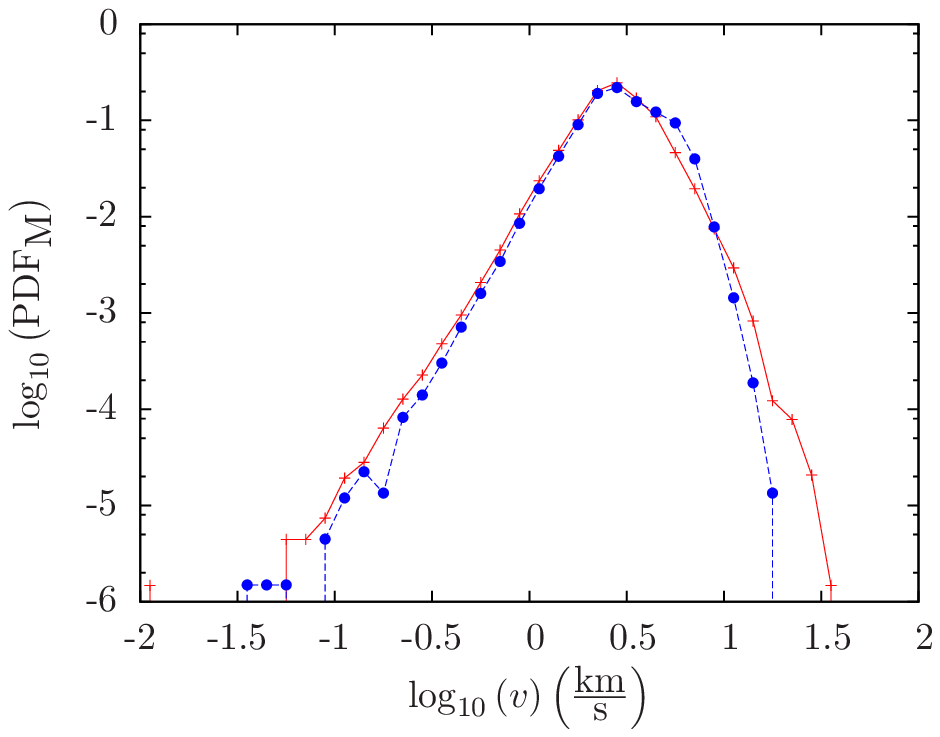}
\\
{Run E}
\label{fig:VelPDFRUNE191}
\end{minipage}
\hfill
\begin{minipage}[c]{0.3\textwidth}
\centering
\includegraphics[width=\textwidth]{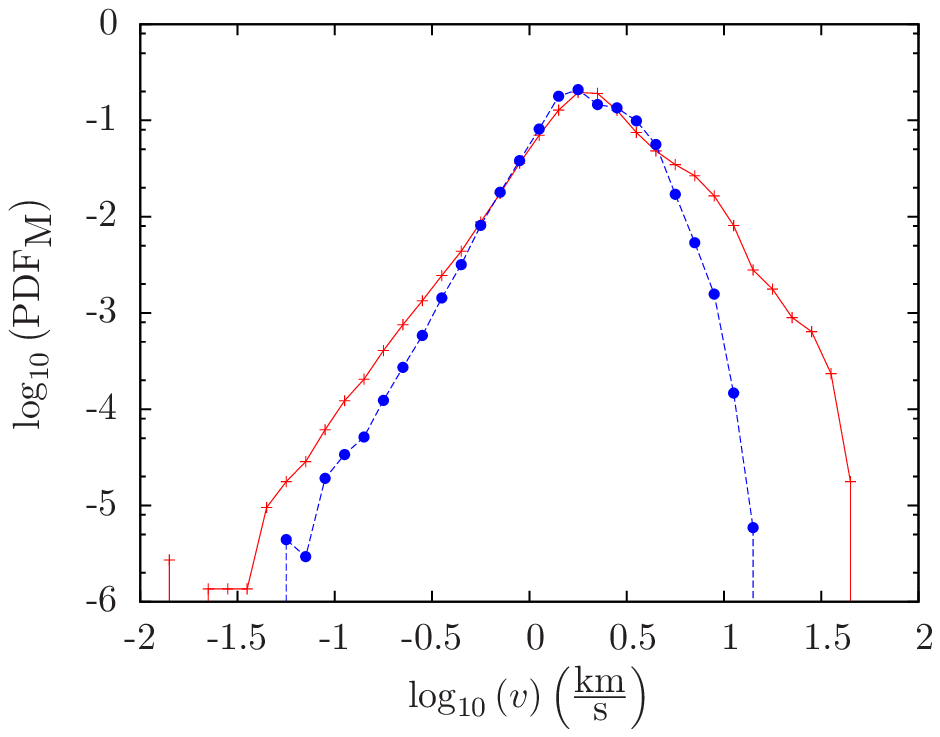}
\\
{Run J}
\label{fig:VelPDFRUNJ113}
\end{minipage}
\hfill
\begin{minipage}[c]{0.3\textwidth}
\centering
\includegraphics[width=\textwidth]{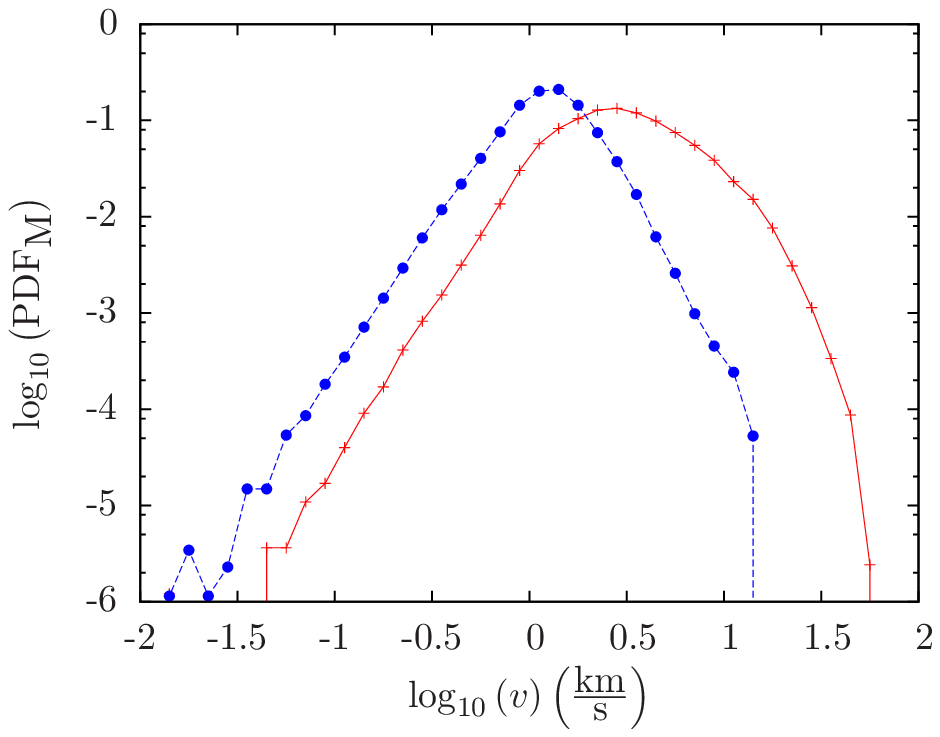}
\\
{Run I}
\label{fig:VelPDFRUNI180}
\end{minipage}\vspace{5pt}
\caption{Density PDFs (upper row) and velocity PDFs (lower row) for Run~E ($t=7.74\text{Myr}$), Run~J ($t=3.49\text{Myr}$) and Run~I ($t=7.58\text{Myr}$).  The runs including photoionising feedback are represented by the red crosses and solid lines and the control run simulations by the blue dots and dashed lines. The ionised particles in the respective runs were excluded.}
\label{fig:DensityVelocity2PDF}
\end{figure*}

The upper row in Figure \ref{fig:DensityVelocity2PDF} shows the density probability distribution functions (PDFs), the lower row the velocity PDFs of Run E, J and I at the end of the simulations. The control runs (without ionisation) are marked by the blue dots and dashed lines, the PDFs of the ones including photoionisation by the red crosses and solid lines. The ionised particles were excluded from the analysis since we are interested in turbulence in the cold star-forming gas. This has the consequence that the PDFs of the control runs range to a lower minimum density in all three cases. Overall we note that the density PDFs do not show a significantly larger width in the cold regions. Following the theoretical model by \citet{Vazquez94} and \citet{PassotVazquezSemadeni1998} the PDFs do not suggest that photoionisation drives turbulent motions. However, the maximum density of the PDFs is set by the sink particle formation criteria. Theoretically there is the possibility for much higher densities than the sink particle density. In practice most dense regions collapse and form sink particles at the threshold density. This sets an effective upper cut-off on the PDFs. The much larger extent towards low-density regions is excluded by neglecting the hot gas.

In Run E, where feedback has the least impact on the morphology of the cloud, the PDF of the ionised run goes up to higher densities than in the control run, but only for a small number of particles $(\sim 100)$. 
In Run E, the sink particle formation density is lower and the sink particle radii are bigger - as stated in Section~\ref{sec:MethodSimulations}, they represent small clusters in Run E, instead of stars. In Runs I and J, the sink formation density is about $7\times 10^{7} \text{cm}^{-3}$, in Run E the value is lower, namely $4\times 10^{5} \text{cm}^{-3}$ (see Section~\ref{subsec:Simulations}). The gas densities in Run E exceed the sink particle formation density because the density threshold is only a necessary but not sufficient formation criterion. If in addition the dense region under consideration is also bound then the accreting sink particle is formed. \cite{BateEtAl1995} and \cite{FederrathEtAl2010_2} introduce and explain the importance of additional sink particle formation checks.

The peak of the PDF is, in all three cases, at higher densities for the ionised runs than for the control runs. The O-stars are formed in the regions with the highest density gas, so feedback of these stars first destroys this material, thus lowering the PDF at the high density end. On the other hand, photoionisation increases the density of the gas at the boundary of the hot bubble, so the PDF is shifted and the \textit{average} density increases. In Runs I and J, feedback does not seem to effectively create regions of very dense gas. The large-scale structure of the clouds approximately follows an $r^{-2}$ density profile. In simulations where feedback-driven bubbles come to occupy large fractions of the cloud volume, although the mass of swept-up material increases as the bubbles expand, its surface and volume density decline approximately as $1/r$ \citep{DaleEtAl2013wind}. In Run E, the bubbles never exit the denser core of the cloud, so this is less evident.

The lower row of Figure \ref{fig:DensityVelocity2PDF} shows the velocity PDFs. In Run E (left panel) the PDFs of the control run and the ionised run are almost identical. There is only a small deviation for very high $v$, where the control run ranges to $\log \left(v/ \text{km}\hspace{2pt} \text{s}^{-1}\right) \approx 1.25$ and the ionised run to $\log \left(v/ \text{km} \hspace{2pt}\text{s}^{-1}\right) \approx 1.5$. This behaviour is more distinct in Run J and Run I (middle and left panel, respectively), where the highest velocities are almost one order of magnitude higher in the case with photoionisation. These particles with high velocities are found in the opposing extreme regimes in the simulation, i.e. in the collapsing, high density structures, as well as in the expanding regions at large radii. The clouds are not confined, but allowed to expand freely. Especially in Run I and J, photoionisation creates HII bubbles that fill up a large fraction of the cloud volume and does a substantial amount of damage to the cloud, thus leading to high velocities in the outer regions.

Arbitrarily high velocities for the gas particle are excluded by the choice of sink particle parameters. The gas which is accreted onto sink particles reaches maximum free-fall velocities of the order of a few $\mathrm{km}~\mathrm{s}^{-1}$, comparable to the velocity dispersion in the cloud.

\subsection{Effect on final snapshot}
\label{subsec:Final}
In the following section, we study the velocity structure functions in the control and ionised run, respectively. The structure functions were plotted using the data of the final step of the simulations, i.e. they correspond to the column density maps  and PDFs presented in the previous figures. Figures \ref{fig:strfctRUNE191}, \ref{fig:strfctRUNJ113} and \ref{fig:strfctRUNI180} show log-log-plots of the velocity structure functions $S(dr)$ of the respective clouds, the colour-coding is the same as before. The Kolmogorov-type structure function with a slope of $2/3$ is given by the black dotted line in each plot for comparison. In the three figures, the intercept of this line varies, as it is used to illustrate the power law behaviour of the structure functions in the respective run and we are locally making a comparison of the slopes. Comparing the structure functions of Run E, J and I, we note a changing range of the y-axis. Also, as the clouds have different initial radii, they reach different sizes when the simulations are stopped. This depends of course also on the escape velocity of the respective cloud and on the resulting impact of feedback on the morphology. 

In all three figures, $S(dr)$ increases strongly at the very large scales, in the ionised run more than in the control run. This is due to the fact that the clouds are not confined, but expand freely. The structure functions in the ionised cases reach higher values than those of the control run, a behaviour which is most likely also due to the tendency of photoionisation to unbind the clouds and to drive them apart. At the very small scales, there are only few particles in the bins when calculating the structure functions, so the results on scales up to $\log \left( dr / \text{pc} \right) \approx -1$ should be treated with caution. It is possible that some interparticle separations are less than the smoothing lengths of either particle, in which case the velocity differences between the two particles may not be meaningful. However, we found that excluding such pairings had negligible effects on our structure functions. 

We will now describe the behaviour of the structure functions for the three runs in detail.
\begin{enumerate}
\item {\textbf{Run E:} Figure~\ref{fig:strfctRUNE191} shows the velocity structure function of Run E, where photoionisation produces only small HII bubbles. In the control run, $S(dr)$ has a dip at scales of $\log\left( dr / \text{pc} \right) \approx 1.5$. This is an indication of energy being transported in a turbulent energy cascade (a relict from the initial seed of turbulence) from large scales to the smaller scales and energy being lost in shocks. Energy is not replenished at the large scales and therefore $S(dr)$ decreases. On the other hand, there is an increase at the smaller scales to which energy is transported. 

The situation is different in the ionised run: the dip that is present in the control run is much less distinct here. Also, the structure function shows a power law behaviour with a slope of ${2}/{3}$ over a range from $\log\left( dr / \text{pc} \right) \approx -0.6$ to $0.9$.  The fact that $S(dr)$ retains its power law shape over a large range of scales and regains it at $\log \left( dr / \text{pc} \right) \approx 1.5$ (described in detail in Section \ref{subsec:timeevol}) is a sign of turbulence being driven on scales of about $10-20 \text{pc}$ (corresponding to the size of the HII bubbles) and leading to a turbulent energy cascade at intermediate values of $dr$.}\\

\begin{figure}
\centering
\includegraphics[width=\columnwidth]{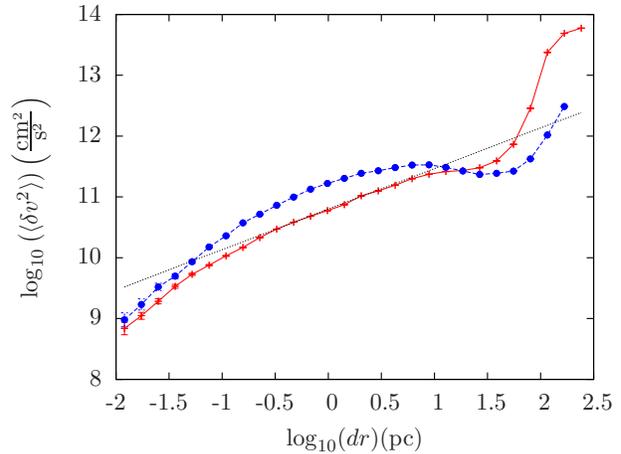}
\caption{Structure function of Run E for $t=7.74\text{Myr}$ (final snapshot): ionised run (red crosses, solid line), control run (blue dots, dashed line), Kolmogorov-type velocity field (black dotted line)}
\label{fig:strfctRUNE191}
\end{figure}

\item \textbf{Run J:} Signs of the decay of the initial Kolmogorov-like turbulent energy cascade and energy being transported from the large to the smaller scales without being replenished, are more pronounced in Run J. This can be seen in Figure \ref{fig:strfctRUNJ113}. Here, the dip in the structure function of the control run simulations is even more prominent than in Run~E. Due to the different initial sizes and therefore different evolution of these two clouds, this is at smaller scales in Run~J, namely around  $\log \left( dr / \text{pc} \right) \approx 1$. Here again, energy has been transported to smaller scales, but turbulence is not replenished at large scales. The increase of  $\log \left( \langle \delta v^2 \rangle \right)$ towards smaller scales, i.e. between $\log \left( dr / \text{pc} \right) \approx -1.5$ and $0.5$ is caused by gravitational collapse.

This is completely different in the case including photoionising feedback, where the signs of gravitational collapse and draining of energy are not present. We find a structure function that approximately follows a power law slope of ${2}/{3}$ from $dr \approx 0.05\text{pc}$ up to scales of about $10\text{pc}$. The power law behaviour over a large range of length
scales can be interpreted as turbulence being driven on large scales and the development of a Kolmogorov-type cascade. An alternative interpretation is that turbulence is being driven over a large range of scales. We will focus more on the interpretation of the differences between the control and ionised run when studying the time evolution of the respective structure functions in Section \ref{subsec:timeevol}.\\

\begin{figure}
\centering
\includegraphics[width=\columnwidth]{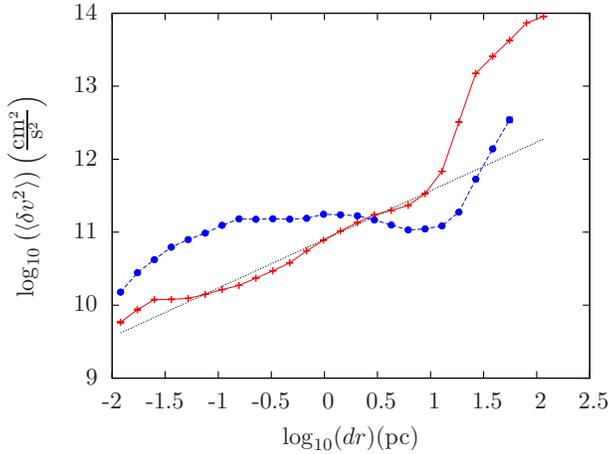}
\caption{Structure function of Run J for $t=3.49\text{Myr}$ (final snapshot): ionised run (red crosses, solid line), control run (blue dots, dashed line), Kolmogorov-type velocity field (black dotted line)}
\label{fig:strfctRUNJ113}
\end{figure}

\item \textbf{Run I:} The features of the structure functions and their respective interpretation are even more prominent in Figure \ref{fig:strfctRUNI180}, which shows $S(dr)$ of Run I. This is the cloud where photoionisation clears very large HII bubbles and unbinds more than half of the cloud. In the control run, we see again a dip in the structure function, here at scales of  $dr\approx 10\text{pc}$. There is also an increase in $\log\left( \langle \delta v^2 \rangle \right)$ at smaller $dr$ in comparison to the ionised run, this is again a sign of gravitational collapse.
We saw in Section \ref{sec:PDFs} that in the ionised run, feedback destroys parts of the very dense gas and therefore of the potentially collapsing structures. In the control run, this is not the case, leading to these relatively large velocity differences at the very small scales.\\
On the other hand, when examining $S(dr)$ for the ionised run, we find an almost perfect power law shape with a slope of ${2}/{3}$ over a very large range of length scales, from $dr \approx 0.05\text{pc}$ up to about $10\text{pc}$. The prominent increase at larger scales is again because the cloud is not confined. We interpret the fact that the structure function of the ionised case of Run I has this power law behaviour as again a sign of turbulence being driven. We will analyse the time evolution of the structure functions in Section~\ref{subsec:timeevol}. We will see that $S(dr)$ in the ionised run of cloud I shows first signs of gravitational collapse and energy being transported to smaller scales without being replenished before the first stars are born. But, once photoionisation starts, it regains its power law shape.
\end{enumerate}

\begin{figure}
\centering
\includegraphics[width=\columnwidth]{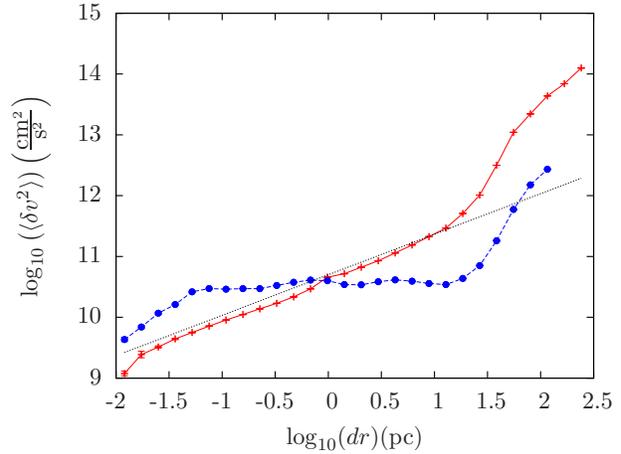}
\caption{Structure function of Run I for $t=7.58\text{Myr}$ (final snapshot): ionised run (red crosses, solid line), control run (blue dots, dashed line), Kolmogorov-type velocity field (black dotted line)}
\label{fig:strfctRUNI180}
\end{figure}

\subsection{Time evolution of structure functions}
\label{subsec:timeevol}
In this section, we illustrate the time evolution of the structure functions in the respective clouds for both the control and the ionised run. In all three cases, $S(dr)$ of the control run is represented by dots and dashed lines, those of the ionised run by crosses and solid thin lines. The black dotted line is again the Kolmogorov-type power law structure function with slope ${2}/{3}$. The red and black thick solid lines and triangles are structure functions that display the very early behaviour of both the control and the ionised run before the first stars are born and photoionisation feedback is switched on. Different colours represent later times, which differ amongst the three clouds. The final stages of the simulations are given in magenta, corresponding to a time shortly before the first supernovae are due to detonate and they are therefore the same as discussed in the previous section. As we are focussing on O-type stars in the simulations and their time on the main sequence is approximately $3\text{Myr}$, this is about the time the simulations will run after ionisation has been enabled. We will describe the time evolution for each cloud separately.
\begin{enumerate}
\item \textbf{Run E:} The black, thick solid line in Figure \ref{fig:timeevolRUNE} is the structure function for a time of $t=5.37\text{Myr}$. It is the same in the ionised and control run as at that time, no stars have been born yet and there is therefore no impact of photoionising feedback. One can still see remnants of the initial power law, especially for the smaller scales. There are early signs of gravitational collapse and the dissipation of energy without replenishment. The green lines represent a time of $t=5.79\text{Myr}$ when photoionisation has already started to have an effect; the dots are for the control run, the crosses for the ionised one. At this stage, the differences between the structure functions are not yet very distinct. This changes for a time of $t=6.68\text{Myr}$ (blue), where the control run structure function evolves a dip around $\log \left( dr / \text{pc} \right) \approx 1.5$ that is much less prominent in the ionised run. Furthermore, the cloud is expanding to larger scales in the run with photoionisation, a behaviour that becomes even more pronounced at later times. The very final stages of the evolution ($t=7.74\text{Myr}$) are displayed in magenta. These structure functions are the same as those described in Section \ref{subsec:Final} and display the gravitational collapse and dissipation of energy in shocks and in the energy cascade in the control run and the driving of turbulence in the ionised run. Overall, the structure functions do not evolve very significantly in comparison to the initial power law, but they do show at least some trend. This is also due to the fact that the created HII bubbles are relatively small in comparison to the size of the whole cloud.\\

\begin{figure}
\centering
\includegraphics[width=\columnwidth]{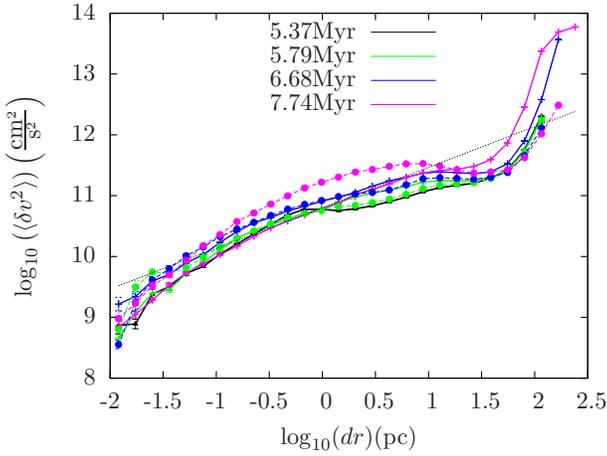}
\caption{Time evolution of the structure functions of Run E: ionised run (crosses, solid thin lines), control run (dots, dashed lines), Kolmogorov-type velocity field (black dotted line). The triangles and thick solid line (black) are for times before photoionising feedback is switched on,  i.e. they are the same in the control and ionised runs. The colours represent different times (black: $t=5.37\text{Myr}$, green: $t=5.79\text{Myr}$, blue: $t=6.68\text{Myr}$, magenta: $t=7.74\text{Myr}$).}
\label{fig:timeevolRUNE}
\end{figure}

\item \textbf{Run J:} $S(dr)$ of Run J in Figure \ref{fig:timeevolRUNJ} evolves substantially in time. The black line and triangles represent a very early time in the evolution of the cloud ($t=0.75\text{Myr}$), where the initial turbulence and hence the power law shape is visible over a large range of length scales. We fit the slope of the structure function over this range and obtain a value of $0.94$, very close to the expected value of $1.00$ for the initially-imposed Burgers velocity field. As time passes, we see the first signs of gravitational collapse at $t=2.10\text{Myr}$ (red line), namely an increase of $\log\left( \langle \delta v^2 \rangle \right)$ at the small scales. These structure functions are the same for the ionised and the control run as photoionisation is only switched on around $t=2.30\text{Myr}$ with the first massive stars having formed. From that point on, the runs start to evolve very differently: In the control run, energy is transported to the smaller scales leading to an increase of $S(dr)$ when following the evolution from green ($t=2.55\text{Myr}$) and blue ($t=3.00\text{Myr}$) to magenta ($t=3.49\text{Myr}$), which is the final stage of the simulation and therefore the same structure function as in Figure \ref{fig:strfctRUNJ113}. This is different in the ionised run, where we see that the structure function is slowly regaining a Kolmogorov-type power law shape once photoionisation begins to exert its influence. The dip at large scales is "refilled" and a turbulent energy cascade regained, which we interpret as a sign of turbulence being driven at the large scales. Photoionisation feedback thus puts energy back into the system which then serves as a driver for turbulence. Then, the turbulent cascade transports energy to smaller $dr$. The formed HII bubbles are expanding radially, i.e. acting as a compressive force, but due to shear this also leads to turbulent eddies and therefore rotational modes in the velocity field, as we show later in Section~\ref{subsec:Modes}.\\

\begin{figure}
\centering
\includegraphics[width=\columnwidth]{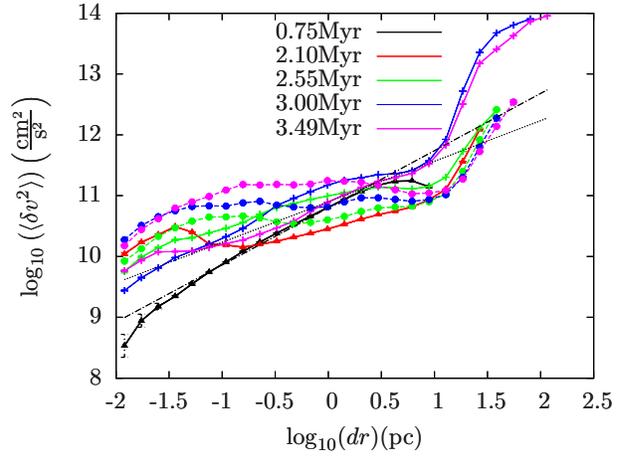}
\caption{Time evolution of the structure functions of Run J: ionised run (crosses, solid thin lines), control run (dots, dashed lines), Kolmogorov-type velocity field (black dotted line). The triangles and thick solid lines (black, red) are again for times before photoionising feedback is switched on. The colours represent different times (black: $t=0.75\text{Myr}$, red: $t=2.10\text{Myr}$, green: $t=2.55\text{Myr}$, blue: $t=3.00\text{Myr}$, magenta: $t=3.49\text{Myr}$). The dash-dotted line is a fit to the power law portion of the structure function at the earliest timestep and has a slope of $0.94$.}
\label{fig:timeevolRUNJ}
\end{figure}

\begin{figure}
	\centering
	\includegraphics[width=\columnwidth]{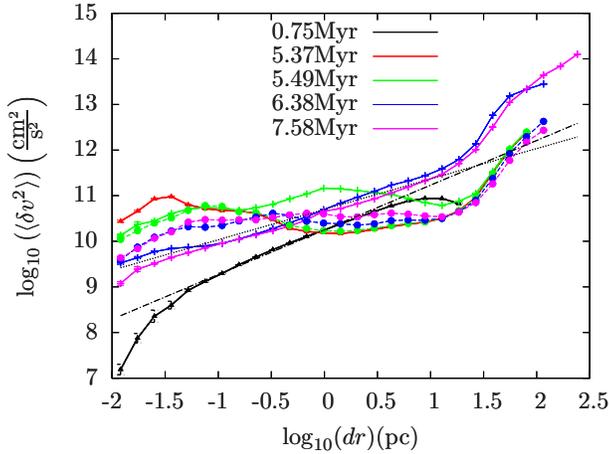}
	\caption{Time evolution of the structure functions of Run I:  ionised run (crosses, solid thin lines), control run (dots, dashed lines), Kolmogorov-type velocity field (black dotted line). The triangles and thick solid lines (black, red) are for times before photoionising feedback is switched on. The colours represent different times (black: $t=0.75\text{Myr}$, red: $t=5.37\text{Myr}$, green: $t=5.49\text{Myr}$, blue $t=6.38\text{Myr}$, magenta: $t=7.58\text{Myr}$). The dash-dotted line is a fit to the power law portion of the structure function at the earliest timestep and has a slope of $0.98$.}
	\label{fig:timeevolRUNI}
\end{figure}

\item \textbf{Run I:} The structure functions of Run I start off with the initial power law shape (black thick line, triangles) at $t=0.75\text{Myr}$ just as in the aforementioned clouds. We again fit the slope of the power law portion of the structure function at this time, yielding a value of 0.98 which is again very close to the analytic structure function slope of 1.00 appropriate for Burgers turbulence. We are therefore confident that our analysis is recovering the statistical characteristics of the velocity fields accurately.
Similarly to Run J, we can see an increase of $S(dr)$ at small scales and a decrease at scales between $\log \left( dr / \text{pc} \right) \approx -0.25$ and $1.25$. This is illustrated by the red line at a time of $t=5.37\text{Myr}$, i.e. before the first stars are born and feedback starts to impact the surrounding cloud about $0.1\text{Myr}$ later (green lines). In analogy to Run J, the control run structure functions keep this shape as there is no mechanism driving turbulence. By contrast, photoionisation feedback quickly leads to an energy input at relatively large scales and makes the structure function regain its power law shape over a large range of length scales for times after $t=6.38\text{Myr}$. This shape is retained until the end of the simulation 
 at $t=7.58\text{Myr}$, which we again interpret as turbulence being driven, leading to a Kolmogorov-type turbulent energy cascade as described above.\end{enumerate}

\subsection{Rotational and compressive turbulent modes}
\label{subsec:Modes}
\begin{figure*}
  \begin{minipage}{\textwidth}
    \centering
    \includegraphics[width=8cm]{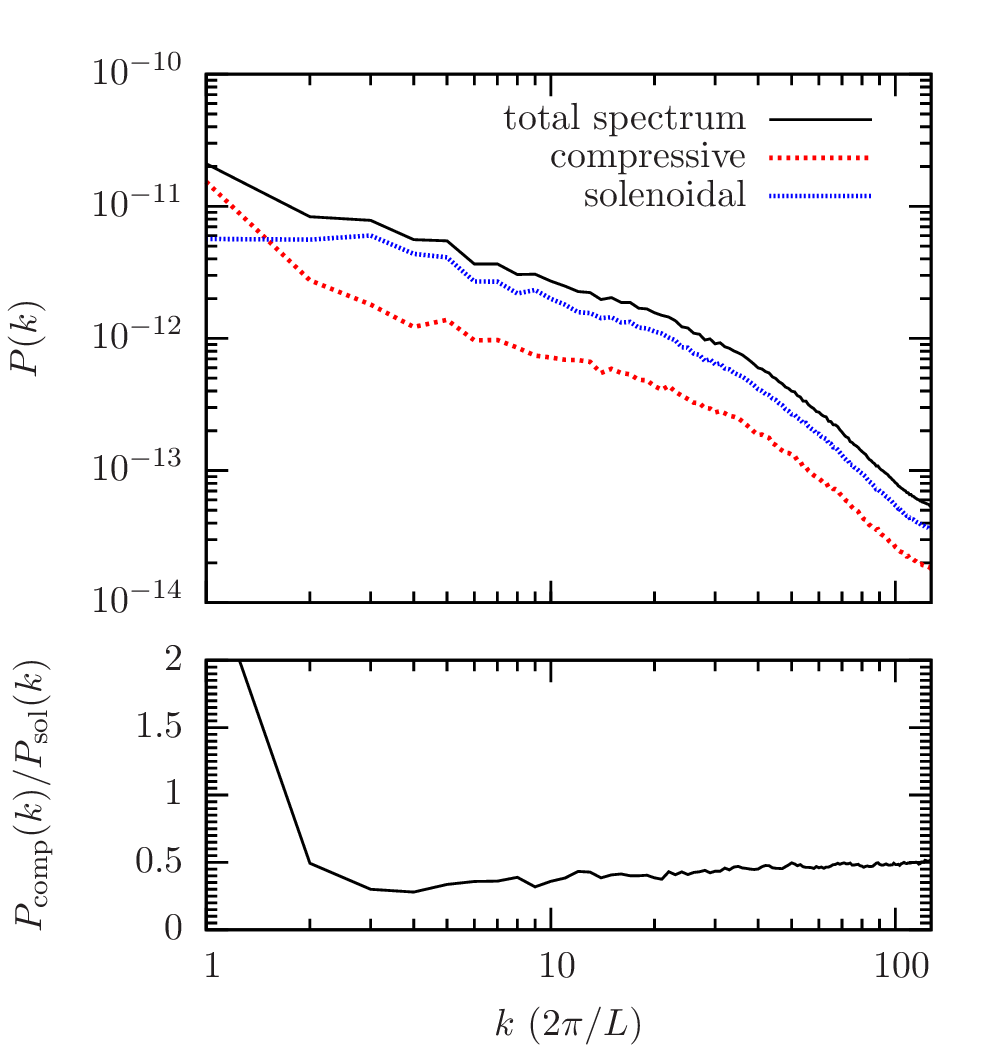}
    \includegraphics[width=8cm]{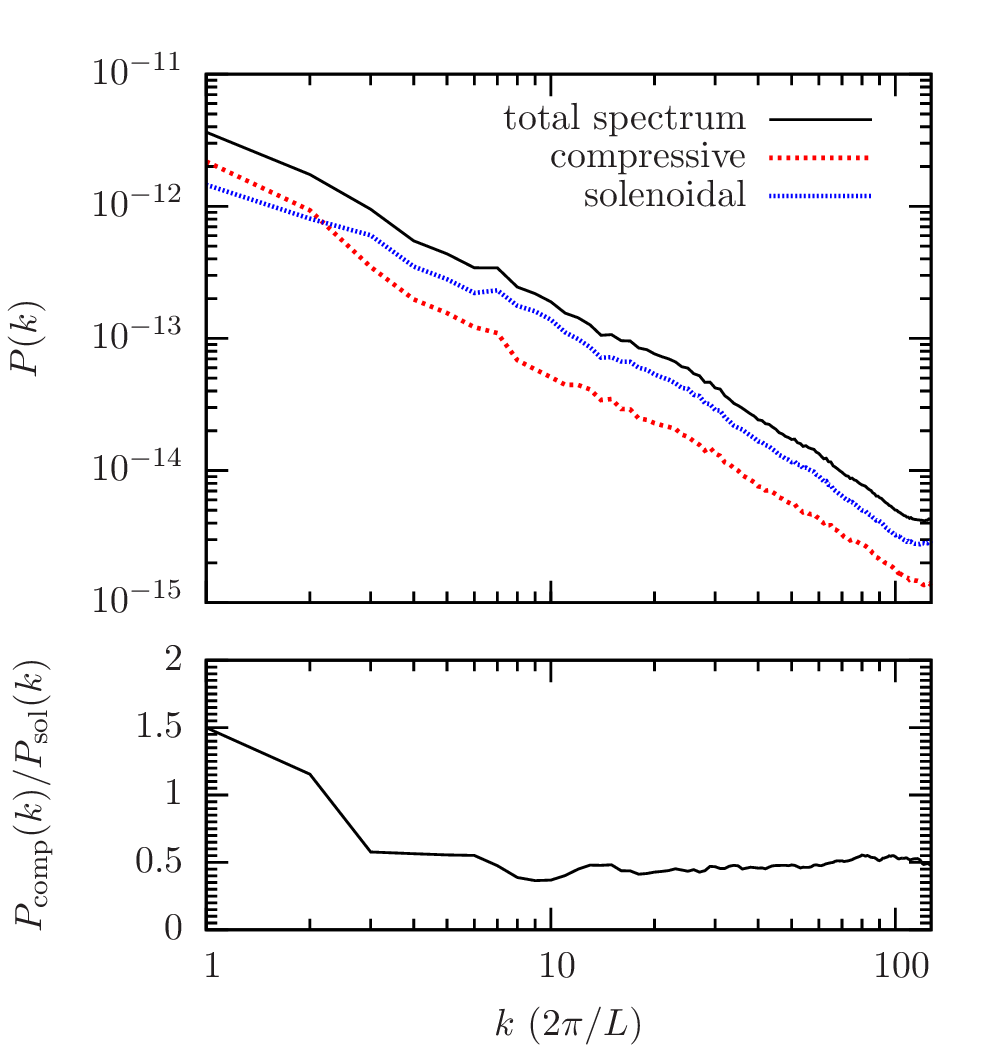}
  \end{minipage}
  \caption{Mass-weighted velocity spectra for the ionisation Run E (left) and the control Run E (right).}
  \label{fig:spec-runE}
\end{figure*}
\begin{figure*}
  \begin{minipage}{\textwidth}
    \centering
    \includegraphics[width=8cm]{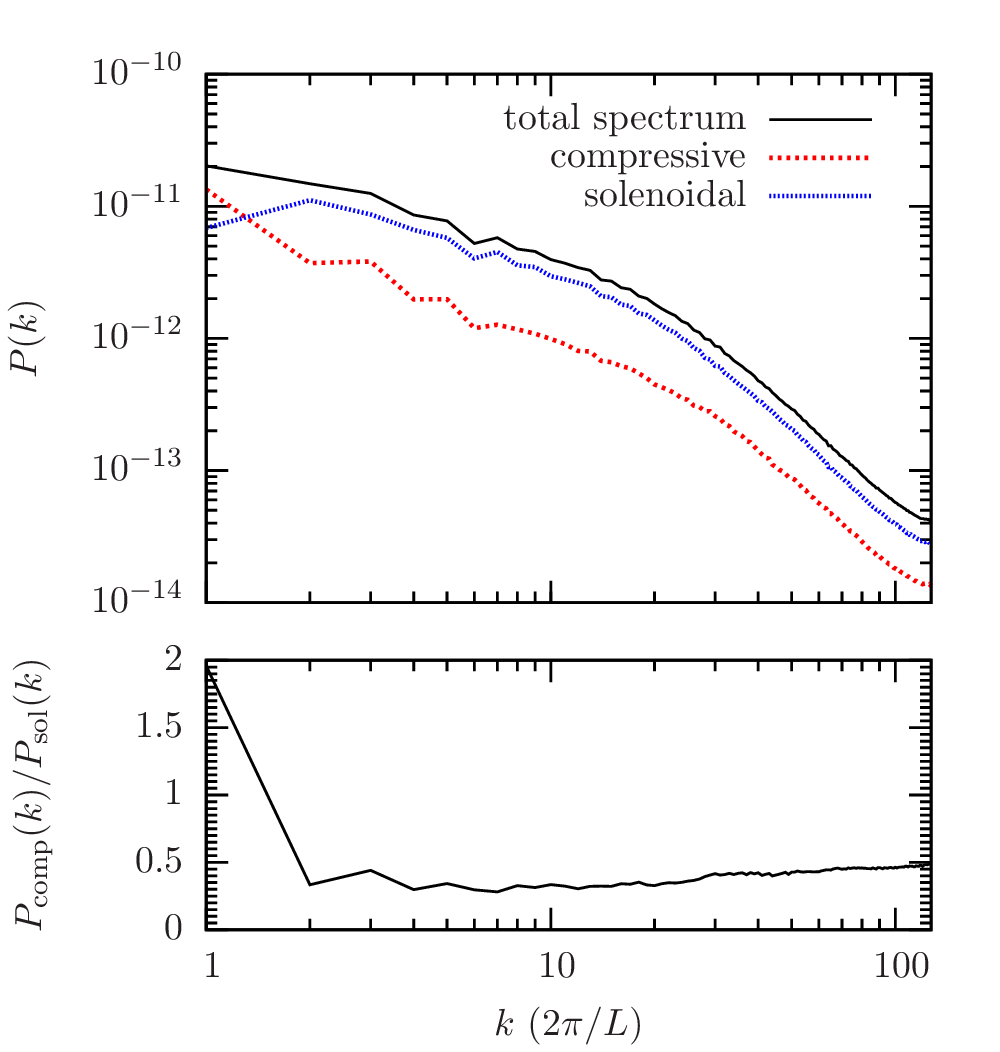}
    \includegraphics[width=8cm]{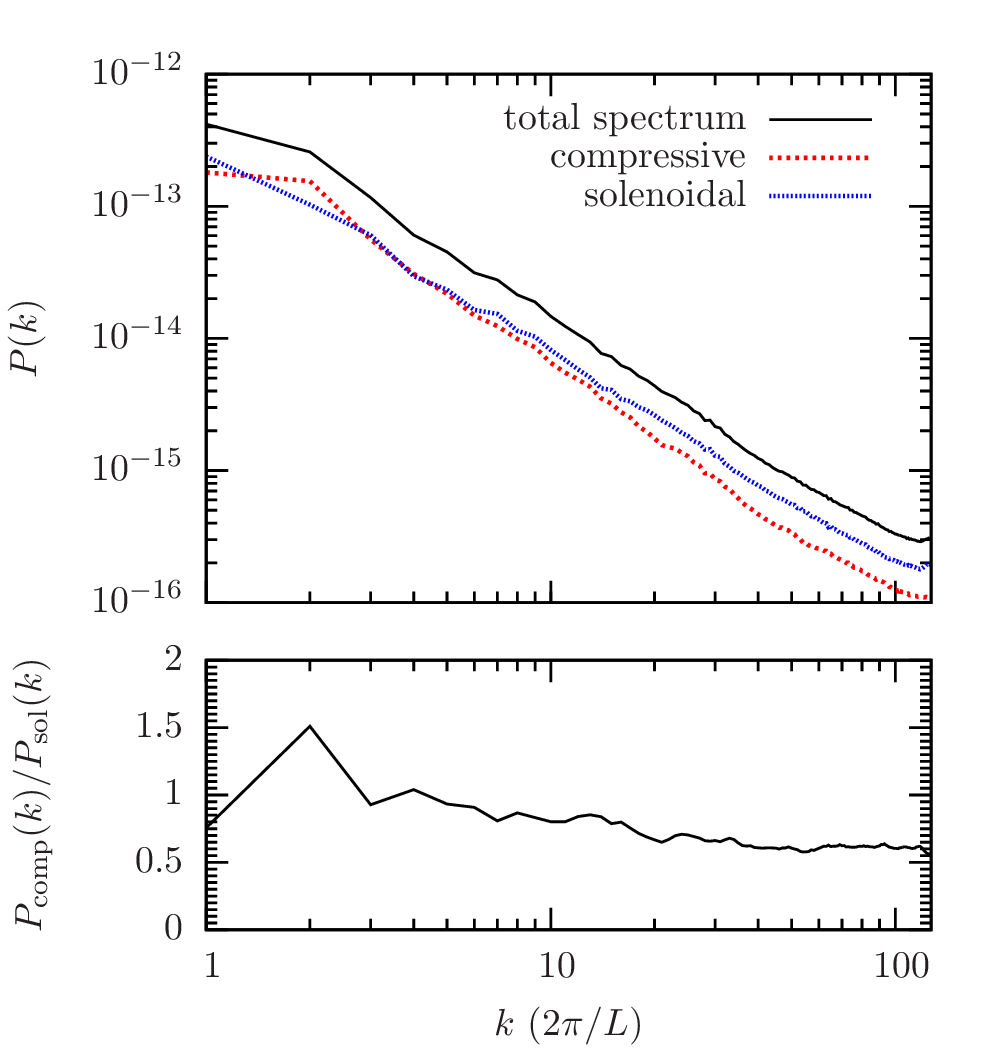}
  \end{minipage}
  \caption{Mass-weighted velocity spectra for the ionisation Run I (left) and the control Run I (right).}
  \label{fig:spec-runI}
\end{figure*}

In this section we focus on the Fourier velocity spectra at the end of the simulations as discussed in Section~\ref{subsec:Spectra}. In order to perform a Fourier transform we map the SPH particles onto a uniform grid with a resolution of $256^3$. A standard SPH sum is computed at the centre of each grid cell, using all SPH particles overlapping at that point, to assign kernel-weighted densities (and therefore masses) and velocities to the cell. The spectral analysis is performed for four simulations, Run~E and Run~I with and without ionisation feedback. We choose Run~E and Run~I as these are two extreme cases in terms of impact of photoionisation feedback. The grid covers the central area of the simulated cloud, which fits in a box with an edge length of $50\,\mathrm{pc}$ for Run~E and $30\,\mathrm{pc}$ for Run~I.
We note that the resolution of the interpolated grid is relatively low compared to the actual resolution of the SPH simulation in regions where the gas density is highest (and therefore where the particle smoothing lengths are smallest).
The small-scale modes of the Fourier transform are thus not affected by effects of the resolution limit in the simulations. We focus on the mass-weighted spectra as we are mainly interested in the effect of photoionisation feedback on the dense, cold star-forming gas. 

Figures~\ref{fig:spec-runE} and \ref{fig:spec-runI} show the spectra for Run~E and Run~I. We show the mass-weighted spectra for the ionisation run (left) and the control run (right). Within each panel the decomposed spectra and the ratio of compressive to solenoidal component are shown. The control runs have spectra with a functional form close to a power law over the entire range. 
The runs including ionising radiation show overall significantly higher amplitudes indicating the additional dynamical impact of the ionisation. In addition, they show deviations from a power law with a flatter slope in the range of $k\lesssim30$. We refrain from performing a detailed analysis of the spectra like determining a spectral slope because general assumptions of fully developed turbulence are not fulfilled. Nonetheless we note that in the case of ionisation the energy input on scales of $k\lesssim30$ has a significant impact on the power spectra.

On scales of the box size, i.e. at $k=1$, the compressive modes dominate the spectrum indicating gravitational contraction (Run~E) or pressure-driven expansion (Run~I). On these scales there is no time for turbulence to develop a composition of modes with the expected ratio. The average crossing time of particles at the largest radii turns out to be larger than the total simulation time in all cases.
The mass-weighted spectra with ionisation feedback show a well-balanced composition of modes with a ratio of $P_\mathrm{comp}/P_\mathrm{sol}\approx0.5$ over a large range of scales. This seems to be consistent with the results of \cite{FederrathEtAl2011}, where the authors study this ratio for a collapsing, self-gravitating cloud. This value of $P_\mathrm{comp}/P_\mathrm{sol}$ is not reached in our control Run~I, which is collapsing very fast. The strong compressive driving due to self-gravity thus dominates the modes and does not allow for depositing energy in rotational modes.

The mass-weighted spectra pronounce the motions in dense regions, which indicate that the enhanced motions due to ionisation have the signatures of fully developed turbulence with roughly the expected ratio of modes. The spectra suggest that radiation drives turbulent motions on scales of $k\lesssim30$ corresponding to $l\gtrsim1\,\mathrm{pc}$. The composition of modes in the runs including ionisation suggest that the energy transfer from the hot bubbles into the surrounding medium is not predominantly a fast compressively driven energy input. Instead, radiation primarily heats the gas and allows for an equi-partitioned dynamical evolution of turbulent motions in dense regions.

\subsection{Initially unbound clouds}
\label{subsec:Unbound}
In a subsequent paper, \cite{Daleetal2012unbound} examined the effect of photoionisation on initially globally \emph{unbound} turbulent clouds. The simulation setup was identical to that described above, except that the normalisation of the turbulent velocity fields was changed to give the clouds initial virial ratios of 2.3 instead of 0.7 (corresponding to turbulent Mach numbers initially a factor of 1.8 times larger for a cloud of a given mass and radius). Run UQ from \cite{Daleetal2012unbound}, for example, has the same initial mass and radius as Run J but its initial turbulent velocity is increased to 5.4 km s$^{-1}$ from Run J's 3.0 km s$^{-1}$.\\
\indent The dynamical effect of ionisation feedback on the unbound clouds was found to be similar to the effect on the bound clouds of \cite{Dale:2012tp}. The principal factor determining how much material was unbound by feedback was the escape velocity of the cloud.\\
\indent We applied the same analysis described above to the evolution of the turbulent structure functions of a selection of these initially unbound clouds, namely Runs UQ, UC, UV, and UZ (see \cite{Daleetal2012unbound} for details of these simulations). For brevity, we do not reproduce the analysis here, but we will simply describe the results.\\
\indent We find that the evolution of their turbulent velocity fields is very similar to the behaviour observed here. As explained above, Run UQ is an unbound analogue of Run J and we find that the evolution of Run UQ's velocity field, as measured by the structure function, is very similar to that of Run J described above. Run UV is an unbound analogue of Run E and the evolution of the velocity fields of these two calculations are again very similar. In general the velocity field of a given unbound cloud behaves in the same  manner to that in the corresponding bound cloud: lower-mass clouds show strong evolution away from a power law structure function in the control simulations, but a power law close to the Kolmogorov slope is restored in the counterpart feedback simulations. In the higher-mass clouds, the evolution of the structure functions in the control simulations is more modest, but departures from the initially-imposed power laws are again reduced by the action of photoionisation.
\subsection{Effects of self-gravity on the velocity field}
\label{sec:Discussion}
\begin{figure}
\centering
\includegraphics[width=\columnwidth]{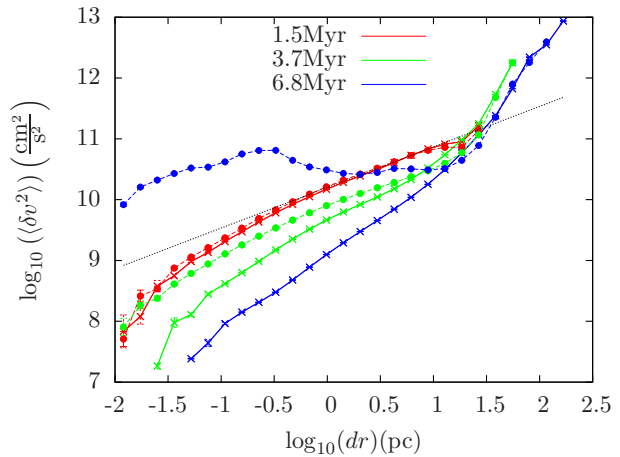}
\caption{Structure functions of Run I at different times, for the control run (dots, dashed lines) and a run without self-gravity (crosses, solid line). Different colours represent different times (red: $t=1.5 \text{Myr}$, green: $t=3.7 \text{Myr}$, blue: $t=6.8\text{Myr}$). The Kolmogorov-type reference line is given by the black dotted line.}
\label{fig:RUNI_nograv}
\end{figure}

In order to test which effect gravity has on its own (as given in the control run) and which features cannot be explained due to it, we perform a run without self-gravity and without any feedback. As gravitational collapse was most pronounced for Run I, we also choose this cloud here. We then compare the structure functions of these two runs in Figure~\ref{fig:RUNI_nograv}. The control run is again given by the dots and dashed lines, the one without gravity by the crosses and solid lines. We have also plotted the Kolmogorov-type power law structure function for comparison. The different colours represent different times in the evolution. 

For relatively early times (red line, $t=1.5\text{Myr}$), both structure functions almost overlap, as gravity has not yet started to dominate in the control run. Both simulations show at this point the almost power law $S(dr)$ from the initial turbulence with which the clouds were seeded. After this point, the evolution starts to differ: The green lines represent times of $t=3.7\text{Myr}$. Both clouds are expanding as they are not artificially confined, but the run with the impact of gravity shows first signs of  turbulence that is decaying at large scales, namely the aforementioned dip at $dr \approx 10\text{pc}$. This is different in the run without self-gravity. Here the cloud is losing energy over a large range of scales (from the very small ones up to about $dr \approx 10 \text{pc}$). This can be explained by the fact that in this cloud no gravitational collapse will take place, i.e. no stars will form, but  the cloud starts to diffuse and becomes almost homogeneous. 

This trend is also visible for the last timestep displayed here, the blue lines at $t=6.8\text{Myr}$. The cloud without gravity has expanded, but has at the same time lost most of its energy on the smaller scales. It has moved away from the initial power law with slope $2/3$ to a much steeper value. We can thus conclude that there is no longer a Kolmogorov-type turbulent energy cascade at work. For the control run we see as before the clear signs of gravitational collapse that led to a strong increase of $S(dr)$ at the small scales. As the structures functions of both clouds are almost overlapping for the very large scales at late times, we can conclude that the evolution at these scales is not dominated by self-gravity. This was different in the ionised run, where the expanding HII bubbles had enormously changed the velocity field also at these $dr$.\\

The effects of self-gravity and star formation on turbulent density and velocity fields were recently studied by \cite{FederrathKlessen2013} using adaptive mesh refinement (AMR) simulations of driven turbulence in a periodic box. They observed that gas-self-gravity (enabled only once turbulence was established) had a minimal effect on the velocity spectra, except in simulations with low turbulent Mach numbers ($\mathcal{M}\approx 3$ - see their Figure 6). They apply turbulent forcing at wavenumbers in the range $1<k<3$ and attribute this result to the most weakly-driven clouds entering a state of global gravitational collapse. This is essentially what is happening in our clouds, which are subject to decaying turbulence and are gravitationally unstable throughout large fractions of their volumes. \cite{FederrathKlessen2013} also find that self-gravity has a pronounced effect on the \emph{density} power spectra, regardless of the strength of the turbulent driving.

\section{Conclusion}
\label{sec:Conclusion}
We have examined the time evolution and especially the structure functions and spectra of the final snapshots for control and ionised runs of GMCs that show a different impact of photoionising feedback on their morphologies.
Our main  conclusions can be summarised as:
\begin{enumerate}
\item The control simulations of the lower-mass clouds I and J rapidly lose their initial power law form, indicating the decay of their turbulent velocity fields. The corresponding simulations including ionisation feedback, however, rapidly recover a power law structure function characteristic of turbulence. 
\item We find that, in the control simulations of Runs E and I, the ratio of power in compressive to solenoidal modes is in general higher than would be expected from well-developed turbulence, particularly on the largest scales. By contrast, in the ionised calculations, this ratio is close to 0.5 over large ranges of wavenumbers, which is again characteristic of turbulence.
\item The velocity field indicating the presence of turbulence is established on very short timescales in the clouds including feedback. These timescales are shorter than the crossing times in the respective runs. Thus photoionisation offers a means of quickly creating a velocity field bearing the typical signs of turbulence in the cold, dense gas.
\end{enumerate}
We found that in Run E, a cloud with relatively little impact of feedback, all structure functions are relatively similar with just the control run $S(dr)$ showing some signs of gravitational collapse. This is shown by the dip in the structure function at large scales and the increase at small scales which is a result of energy cascading from large to small $dr$ without turbulence being replenished. In the ionised run, this behaviour is suppressed and $S(dr)$ keeps its power law shape over many length scales.

In the other two clouds studied, we found more prominent signs of gravitational collapse and transport of energy to smaller scales in the control run. The structure functions of the ionised run start off with an initial power law shape, then show some signs of collapse and draining of energy. Once photoionisation starts, $S(dr)$ approaches a straight line with slope $2/3$ over several orders of magnitude in scale. We interpreted this as turbulence being driven by interacting HII bubbles. This is in good agreement with the results from \cite{Gritschnederetal2009}, who also find that photoionising radiation can be an internal source of driving of turbulence in GMCs.

The analysis of the clouds using power spectra reveals that a significant amount of kinetic energy is injected on scales of the order of $l\gtrsim\mathrm{few}\,\mathrm{pc}$. The decomposition into compressive and solenoidal modes indicates that the energy enhancement in the dense regions due to radiation is in agreement with the statistical ratio of compressive to solenoidal modes expected from well-developed turbulence.
However, the spectra overall do not show a power law behaviour, violating the classical turbulent cascade where the energy is driven at the largest scales. In the simulations including ionising radiation the \emph{thermal} driving scales are $\sim 1-10 \text{pc}$, which shapes the spectra accordingly.


\section{Acknowledgements}
We thank the referee for helpful comments and suggestions that improved the paper.
This research was supported by the DFG cluster of excellence `Origin and Structure of the Universe' (JED, BE). P.G. acknowledges support from the DFG Priority Program 1573 {\em Physics of the Interstellar Medium}.

\label{lastpage}
\bibliographystyle{mn2e}
\bibliography{paper}
\end{document}